\documentclass[preprint,12pt,3p]{elsarticle}
\usepackage[english]{babel}
\usepackage{amssymb}
\usepackage{latexsym,amsmath,amsthm,amsfonts,graphicx,float}
\usepackage{hyperref}
\usepackage{subfig}
\usepackage{color}
\RequirePackage{natbib}
\usepackage{graphicx}
\DeclareGraphicsExtensions{.pdf,.jpg,.png,.jpeg}
\usepackage[utf8]{inputenc}
\usepackage[T1]{fontenc}
\usepackage{lmodern} 
\usepackage{tikz}
\usetikzlibrary{calc}
\usepackage{caption}
\usepackage{multirow}
\usepackage[bottom,flushmargin,hang,multiple]{footmisc}
\usepackage{etoolbox}
\usepackage{amssymb}
\AtBeginEnvironment{quote}{\small}
\usepackage{arydshln}
\usepackage{comment}

\def \vW{\mbox{\boldmath $W$ \unboldmath}\!\!}

\makeatletter     
\newcommand{\distas}[1]{\mathbin{\overset{#1}{\kern\z@\sim}}}%
\newcommand{\distras}[1]{%
  \savebox{\mybox}{\hbox{\kern3pt$\scriptstyle#1$\kern3pt}}%
  \savebox{\mysim}{\hbox{$\sim$}}%
  \mathbin{\overset{#1}{\kern\z@\resizebox{\wd\mybox}{\ht\mysim}{$\sim$}}}%
}
\makeatother

\journal{NeuroImage}

\begin{document}
\bibliographystyle{elsarticle-num}
\begin{frontmatter}

\title{Neuroimaging Feature Extraction using a Neural Network Classifier for Imaging Genetics}


\author[label1,label3]{C\'edric Beaulac\corref{cor2}}
\ead{beaulac.cedric@gmail.com}
\cortext[cor2]{I am the corresponding author.}
\author[label2]{Sidi Wu}
\address[label1]{Department of Mathematics and Statistics, University of Victoria, Victoria, BC.}
\author[label3]{Erin Gibson}
\address[label3]{School of Engineering Science, Simon Fraser University, Burnaby, BC.}
\author[label1]{ Michelle F. Miranda}
\author[label2]{Jiguo Cao}
\address[label2]{Department of Statistics and Actuarial Sciences, Simon Fraser University, Burnaby, BC.}
\author[label1]{Leno Rocha}
\author[label3]{Mirza Faisal Beg}
\author[label1]{Farouk S. Nathoo}
\author{for the Alzheimer's Disease Neuroimaging Initiative\fnref{adni}}

\fntext[adni]{Data used in preparation of this article were obtained from the Alzheimer's Disease Neuroimaging Initiative (ADNI) database (\url{http://adni.loni.usc.edu}).  As such, the investigators within the ADNI contributed to the design and implementation of ADNI and/or provided data but did not participate in analysis or writing of this report. A complete listing of ADNI investigators can be found at: \url{http://adni.loni.usc.edu/wp-content/uploads/how\_to\_apply/ADNI\_Acknowledgement\_List.pdf}}

\begin{abstract}

A major issue in the association of genes to neuroimaging phenotypes is the high dimension of both genetic data and neuroimaging data. In this article, we tackle the latter problem with an eye toward developing solutions that are relevant for disease prediction. Supported by a vast literature on the predictive power of neural networks, our proposed solution uses neural networks to extract from neuroimaging data features that are relevant for predicting Alzheimer's Disease (AD) for subsequent relation to genetics. Our neuroimaging-genetic pipeline is comprised of image processing, neuroimaging feature extraction and genetic association steps. We propose a neural network classifier for extracting neuroimaging features that are related with disease and a multivariate Bayesian group sparse regression model for genetic association. We compare the predictive power of these features to expert selected features and take a closer look at the SNPs identified with the new neuroimaging features. 

\end{abstract}

\end{frontmatter}

\section{Introduction} \label{intr}

Brain imaging genomic studies have great potential for better understanding psychopathology and neurodegenerative disorders. While  high-throughput genotyping technology can determine high-density genetic markers single nucleotide polymorphisms (SNPs), neuroimaging technology provides a great level of detail of brain structure and function \citep{sheng2022}. Various modalities of brain imaging can be used to generate meaningful biological information that can in turn be used to evaluate how genetic variation influence disease and cognition. In Alzheimer's disease (AD), structural modalities such as magnetic resonance imaging (MRI) can detect the presence of neuronal cell loss and gray matter atrophy, both biomarkers of neurodegeneration. Such neuroimaging phenotypes are attractive because they are closer to the biology of genetic function than clinical diagnosis\citep{lindenberg2012}.

\bigskip

Imaging genetics data analysis is a statistically challenging task due to the high dimension of both the neuroimages and genetic data. Further increasing the challenge is the fact that the data can be of multiple forms; neuroimages can be collected in multiple formats, e.g. MRI, ElectroEncephaloGram (EEG), Computerised Tomography (CT), Positron Emission Tomography (PET) using different machines and in different institutions. Consequently, it is important to find a general solution to the dimension problem that is applicable on a wide range of data structure.

\bigskip

We consider studies having an emphasis on exploring the relation between genetic variation and brain imaging from structural modalities such as MRI and consider associated statistical methodology for dimension reduction and genetic variable selection. We focus our effort on the identification of SNPs that are potentially related to disease, for example, AD, with brain imaging endophenotypes which have the potential to provide additional structure related to the underlying etiology of the disease. Existing approaches for such analysis are based on considering the imaging data through a specific set of regions of interest (ROIs) (see, e.g., \cite{wang2012identifyingb}, \cite{wang2012identifying}, \cite{greenlaw2017}, \cite{vounou2010discovering}, \cite{zhu2014bayesian}) or they are based on a full voxelwise analysis with statistical models fit at each voxel (see, e.g., \cite{hibar2011voxelwise}, \cite{stein2010voxelwise}, \cite{ge2012increasing},\cite{ge2015kernel}, \cite{huang2015fvgwas}).

\bigskip

The first approach for statistical analysis in studies of imaging genetics developed brain-wide and genome-wide mass univariate analyses \cite{stein2010voxelwise}. A drawback of this framework is that it ignores linkage disequilibrium and the associated multicollinearity between genetic markers as well as dependence between the components of the imaging phenotype. Hibar et al. \cite{hibar2011voxelwise} employed gene-based dimensionality reduction to avoid collinearity of SNP vectors. Vounou et al. \cite{vounou2010discovering} employed sparse procedures based on reduced-rank regression and while Ge et al.  \cite{ge2015kernel} considered multi-locus interactions and developed kernel machine approaches. A review of methods is provided by Nathoo et al. \cite{nathoo2019review}. 

\bigskip

Bayesian joint modelling combining imaging, genetic and disease data has been considered in \cite{batmanghelich2013joint} and \cite{batmanghelich2016probabilistic}. The proposed joint models use logistic regression to relate disease endpoints to imaging-based features and a second regression relates imaging to genetic markers. Spike-and-slab selection is employed in both regression components of the joint model. Hierarchical models accounting for spatial dependence in the imaging phenotype using Markov random fields have been developed in \cite{stingo2013integrative} and \cite{song2021bayesian}. Zhu et al. \cite{zhu2014bayesian} developed a Bayesian reduced rank regression reducing the dimension of the regression coefficient matrix and incorporating a sparse latent factor representation for the covariance matrix of the imaging data based on a gamma process prior. Kundu et al. \cite{kundu2016semiparametric} proposed a semiparametric conditional graphical model for imaging genetics within the context of functional brain connectivity where a Dirichlet process mixture is used for clustering regression coefficients into a modular structure. Azadeh et al. \cite{azadeh2016integrative} developed a voxelwise Bayesian approach that began by partitioning the brain into ROIs and then fitting multivariate regression models to lower-dimensional projections of the voxel-specific data within each ROI separately and in parallel across ROIs.

\bigskip

We investigate here a new approach for extracting imaging features in either the ROI or the voxelwise setting. Statistical learning approaches for feature construction and dimension reduction have been developed based on a number of approaches including Gaussian Mixture Models (GMM) \cite{Chaddad2015} and Support Vector Machines (SVM) \cite{Wang2021,Shahajad2021}. The ability of neural networks (NNs) to effectively reduce the dimension of large data has been known for some time \cite{hinton2006}. Since then, NNs have been at the foundation of multiple feature extraction models \cite{Pradeep2011,Chen2016,El2020} in image analysis.
The autoencoder (AE) is a commonly used NN model for feature extraction \cite{Wang2014,Wang2016}. It consists of two pieces, an encoder and a decoder. The former compresses the data, embedding it within a lower-dimensional representation, while the latter decompresses this representation to its original dimension. Both of these components are optimized simultaneously so as to reduce the reconstruction error. The encoder and the decoder can take various forms but we will assume both are NNs.

\bigskip

Predicting a diagnosis successfully using NNs is supported by a large literature \cite{islam2018, lu2018, lin2018, duraisamy2019, jain2019,mirabnahrazam2022} that has demonstrated that various modern neural network architectures, such as Convolutional Neural Networks (CNNs) \cite{lecun1989,lecun2015,goodfellow2016}, weighted probabilistic neural networks \cite{kusy2018weighted} and ensembles of deep neural networks \cite{zhang2012ensemble,lecun2015} can achieve extremely high accuracy in the classification of MRI and PET scans.  Specifically within the context of imaging genetics, Ning et al. \cite{ning2018} were among the first to apply NN approaches. Their approach was to train a NN taking both imaging data and genetic markers as inputs to predict a binary disease response (AD diagnosis). 

\bigskip

The first main contribution of this paper is to develop a neural networks classifier (NNC) for extracting image-based features that emphasize disease. Our approach is equivalent to an AE where we replace the decoder with a classifier model. Instead of training the encoder in a way that the lower-dimensional representations can lead to a reconstruction with a small error, we train the encoder in a way that the lower-dimensional representations are well suited for the prediction of disease. Doing so ensures that the lower-dimensional representations, the extracted features, are relevant in predicting the neurological disease of interest. We thus combine the strength of AEs for producing low-dimensional representations with the high predictive accuracy of NNCs to extract features relevant to disease diagnosis.

\bigskip

We demonstrate that it is possible to achieve higher prediction accuracy to classify disease status (AD relative to normal controls (NC)) when using NNC features compared with features based on known AD ROIs. This then provides a more relevant set of phenotypes for SNP selection. In our proposed approach, we take the last hidden layer of a NNC as a set of features automatically extracted by the model in a way that distinguishes diseased patients from control patients. Using these features as an endophenotype allows us to identify SNPs that are simultaneously related to the MRI and are also more relevant for predicting disease than expertly chosen features. 
\bigskip

The second main contribution of this paper is that we propose a three-step imaging genetic pipeline: image processing, feature extraction and finally genetic inference. Therefore, this separates the pieces where we do not require interpretability such as image processing and feature extraction from the pieces where we need interpretability, mostly genetic inference. This separation is beneficial for two reasons. First, it allows us to utilize the increased prediction accuracy of blackbox models for feature extraction. Second, it is easy to modify and improve the three pieces individually, making this pipeline applicable to a wide range of data structures. Consequently, the novelty of our pipeline lies in how we utilized well-established models altogether. 
\bigskip

The rest of the paper proceeds as follows. We introduce our proposed pipeline in Section \ref{Pipeline}. Then, in Section \ref{ourPipe} we discuss the implementation of this pipeline, we introduce our experimental testing setup and present our results for testing with ADNI data. Section \ref{Discussion} concludes with a discussion of our experimental results and possible extensions.

\section{Proposed pipeline} \label{Pipeline}

We separate the image processing, the feature extraction and the genetic association study into three distinct steps. This allows us to use black box models for the feature extraction step without suffering from the drawback caused by the lack of interpretability. We first introduce the concept and then provide a more formal definition of our modelling approach. 

\subsection{Concept}




Based on the premise that neuroimaging data is a better representation of the phenotype of interest than clinical diagnostic, we aim at capturing genetic variations related to disease. Due to the high-dimensionality of neuroimaging, we propose NNs to extract features related to disease while simultaneously reducing data's dimensionality. 

\bigskip

\begin{figure}[H]
\begin{center}
\begin{tikzpicture}[->, semithick, scale=0.7]
  \tikzstyle{latent}=[fill=white,draw=black,text=black,style=rectangle,minimum size=3cm]
  \tikzstyle{var}=[fill=white,draw=white,text=black,style=circle,minimum size=0.5cm]
  \tikzstyle{observed}=[fill=black!30,draw=black,text=black,style=circle,minimum size=0.5cm]

  \node[latent]   (x) at (-8,0)  {Genotype};

\node[latent]   (z) at (0,0)  {Endophenotype};
    
\node[latent]   (y) at (8,0)  {AD Diagnosis};

    \path (x) edge              node {} (z);
    \path (z) edge              node {} (y);

\end{tikzpicture}
\end{center}
\caption{A graphical representation of the assumed generative process.}
\label{fig1}
\end{figure}
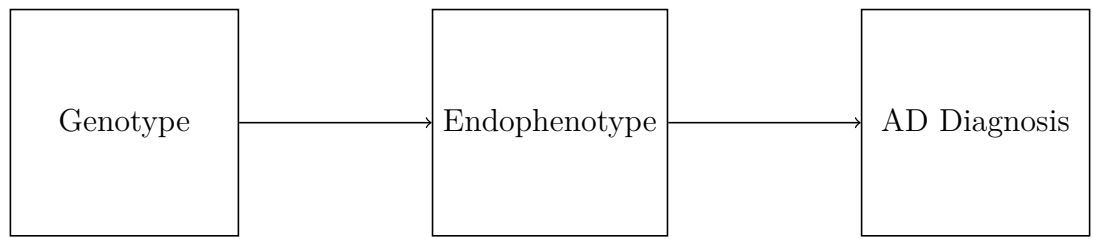

We assume that the natural generation of data follows the premise that genotype is related to brain structure that in turn is related to disease \citep{batmanghelich2016}. Our framework thus reverses the process described in Figure \ref{fig1}, which, while clearly an over-simplification, provides a useful mechanism for thinking about data analysis and SNP selection.

The automated disease-relevant feature extraction is based on training a NN on the imaging data with the disease diagnostic variable as output. The neurons of the second to last layer of this NN prediction function act as the features extracted by the model. Because the NN is optimized to predict disease diagnosis as accurately as possible using the image data, those neurons are in fact the variables constructed from the images that are the most appropriate to predict the disease and are consequently relevant for SNP selection. An alternative, which we make comparisons to in our test analysis are features extracted from known disease regions using expert knowledge. Again, we can make an analogy with an AE. The first few layers of our NNC compress the data to the reduced set of features we wish to extract while the last layer, instead of attempting to reconstruct the input,  replaces the decoder with a loss for disease diagnosis. 

\subsection{Formal definition}

Let $v_{n,m}$ denote voxel $m \in \{1,\dots,M\}$ for subject $n \in \{1,\dots,N\}$ and $\mathbf{v}_n$ denotes the complete imaging data for subject $n$. We identify with $\mathbf{v}^*_n$ the processed image for subject $n$. Here, the processed images may take on different forms but $\mathbf{v}^*_n$ is some standardized image data that the prediction model $f$ takes as input. The processing might only involve image registration in its simplest form or it might involve the extraction of volumetric and cortical thickness statistics using FreeSurfer for instance. Then, $y_n$ is the disease phenotype for subject $n$ which can be binary or multi-class categorical. We further let $g_{n,s}$ denote the genetic variant $s \in \{1,\dots,S\}$ for subject $n$ so that $\mathbf{g}_n$ is the genetic data for subject $n$. 

\bigskip

Let $h$ be the image processing function which takes the images $\mathbf{v}$ as input and outputs $\mathbf{v}^*$. We define as $f$ the classification function which takes $\mathbf{v^*}$, the processed imaging data, as input and outputs $y$, the disease phenotype. We define $f$ as a NN function composed of $L$ layers each identified as ${f_l: l \in(1,L)}$, $f_1$ being the input layer of $f_L$ the output layer. Each layer $l$ may have a different number of neurons $x$, say $K_l$. In our current parametrization, the output layer is a $K_L$-dimensional vector, $\mathbf{o}$, where $o_{n,k} = \hat{P}(y_n = k)$, the predicted probability that subject $n$ belongs to class $k$. After training the neural network $f$, we fit a statistical model, $p$, which has the genetic data $\mathbf{g}$ as explanatory variable and the neurons of the second to last layer of $f$, $f_{L-1}$, as response. 

\bigskip

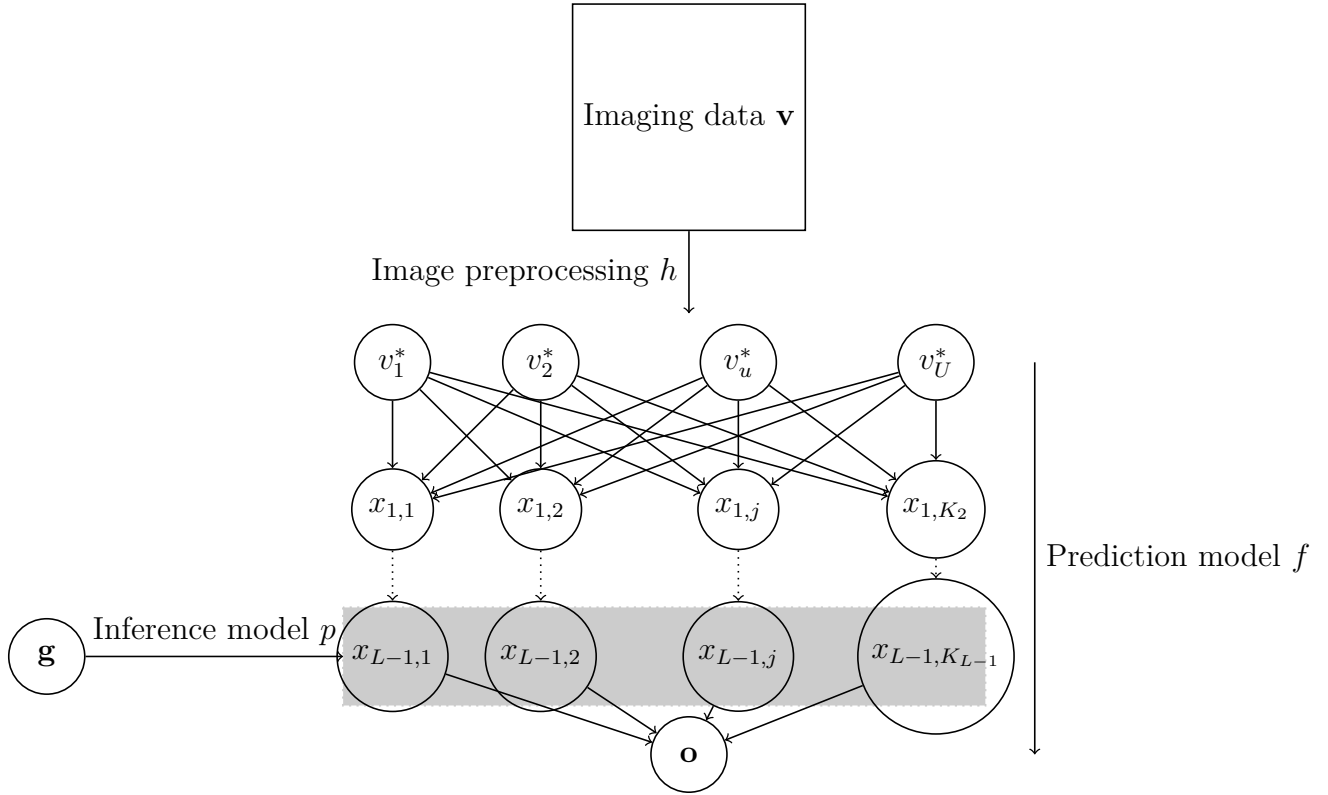
\begin{figure}[h]
\begin{center}
\begin{tikzpicture}[->, semithick, scale=0.65]
  \tikzstyle{lol}=[fill=white,draw=black,text=black,style=rectangle,minimum size=3cm]
  \tikzstyle{var}=[fill=white,draw=black,text=black,style=circle,minimum size=1cm]

  \node[lol]   (v) at (-1,0)  {Imaging data $\mathbf{v}$};

\node[var]   (v1) at (-7,-5)  {$v^*_1$};
\node[var]   (v2) at (-4,-5)  {$v^*_2$};
\node[var]   (v3) at (-0,-5)  {$v^*_u$};
\node[var]   (v4) at (4,-5)  {$v^*_U$};

\node[var]   (x11) at (-7,-8)  {$x_{1,1}$};
\node[var]   (x12) at (-4,-8)  {$x_{1,2}$};
\node[var]   (x1j) at (0,-8)  {$x_{1,j}$};
\node[var]   (x1k) at (4,-8)  {$x_{1,K_2}$};

\node[var]   (xl1) at (-7,-11)  {$x_{L-1,1}$};
\node[var]   (xl2) at (-4,-11)  {$x_{L-1,2}$};
\node[var]   (xlj) at (0,-11)  {$x_{L-1,j}$};
\node[var]   (xlk) at (4,-11)  {$x_{L-1,K_{L-1}}$};

\node[var]   (y) at (-1,-13)  {$\mathbf{o}$};

\node[var]   (g) at (-14,-11)  {$\mathbf{g}$}; 

    \draw[thick,dotted,fill=black,opacity=0.2]     ($(xl1)+(-1,1)$) rectangle ($(xlk)+(1,-1)$);

\path (v) edge              node[left] {Image preprocessing $h$} (-1,-4);

\path (g) edge              node[above] {Inference model $p$} (-8,-11);

\path (6,-5) edge              node[right] {Prediction model $f$} (6,-13);

\path (v1) edge              node {} (x11);
\path (v1) edge              node {} (x12);
\path (v1) edge              node {} (x1j);
\path (v1) edge              node {} (x1k);

\path (v2) edge              node {} (x11);
\path (v2) edge              node {} (x12);
\path (v2) edge              node {} (x1j);
\path (v2) edge              node {} (x1k);

\path (v3) edge              node {} (x11);
\path (v3) edge              node {} (x12);
\path (v3) edge              node {} (x1j);
\path (v3) edge              node {} (x1k);

\path (v4) edge              node {} (x11);
\path (v4) edge              node {} (x12);
\path (v4) edge              node {} (x1j);
\path (v4) edge              node {} (x1k);

\path (x11) edge[dotted]             node {} (xl1);
\path (x12) edge[dotted]               node {} (xl2);
\path (x1j) edge[dotted]               node {} (xlj);
\path (x1k) edge[dotted]              node {} (xlk);

\path (xl1) edge             node {} (y);
\path (xl2) edge               node {} (y);
\path (xlj) edge              node {} (y);
\path (xlk) edge              node {} (y);

\end{tikzpicture}
\end{center}
\caption{A graphical representation of the proposed pipeline.  The prediction model $f$ is depicted as a fully connected NN with $L$ layers though it need not be fully connected.\label{model}}
\end{figure}

A detailed representation of the proposed pipeline is shown in Figure \ref{model}. All components previously described are trained as follows: (i)  process the raw images $\mathbf{v}$, (ii) train a prediction model, $f$, of choice by taking the processed images $\mathbf{v}^*$ as inputs and the diagnosis score as output and (iii) train the inference model $p$ that predicts the features extracted from the prediction model using genetic markers as inputs. The use of a statistical model as our choice of inference model is based on the current availability of interpretable and inference-focused models in the literature.  

\bigskip


The proposed approach can be generalized to include various prediction models such as CNNs taking images as inputs or different NN architectures with inputs being the imaging features extracted from commonly used softwares such as FreeSurfer \cite{dale1999,fischl1999}. Similarly, a wide range of inference models can be used and later combined using Bayesian model averaging techniques that account for model uncertainty at this last stage.

\bigskip

As a proof of concept, in the following test analysis we use disease (AD), MRI and genetic data from the ADNI1 study with FreeSurfer used for image processing, a simple NN for the prediction model and a previously established multivariate group-sparse Bayesian regression model for SNP selection. 

\section{Test application of the proposed pipeline} \label{ourPipe}

For the first implementation of our proposed pipeline, we compare the prediction accuracy of the 56 volumetric and cortical thickness measurements considered in \cite{greenlaw2017}, \cite{shen2010whole}, and \cite{szefer2017multivariate}, which include locations of regions of interest such as the hippocampus, cerebellum and ventricles relevant for AD, with features automatically extracted by our proposed technique. We also compare the SNPs identified given those two sets of phenotype features. 

\bigskip

\textit{Data used in the preparation of this article were obtained from the Alzheimer’s Disease Neuroimaging Initiative (ADNI) database (adni.loni.usc.edu). The ADNI was launched in 2003 as a public-private partnership, led by Principal Investigator Michael W. Weiner, MD. The primary goal of ADNI has been to test whether serial magnetic resonance imaging (MRI), positron emission tomography (PET), other biological markers, and clinical and neuropsychological assessment can be combined to measure the progression of mild cognitive impairment (MCI) and early Alzheimer’s disease (AD).}

\subsection{Cohort of subjects} \label{Cohort}

The cohort of subjects we use in our test application has been previously described by Mirabnahrazam et al. \cite{mirabnahrazam2022}. Briefly, the ADNI1 database has genetic information for 818 subjects. Genotyping information of the ADNI1 subjects was downloaded in PLINK \citep{Chang2015} format from the LONI Image Data Archive (\url{https://ida.loni.usc.edu/}). During the genotyping phase, 620,901 SNPs were obtained on the Illumina Human610-Quad BeadChip platform. Genomic quality control was conducted using the PLINK software and yielded 521,014 SNPs for 570 subjects.  When excluding subjects that had no diagnosis label available we ended up with 543 subjects for our analysis. The diagnosis values we consider for this experiment are NC, MCI and AD.

\bigskip

In summary, we have a cohort of 543 subjects with 145 NC, 256 MCI and 142 AD. We have T1-weighted baseline MRI scans for every subjects as well as 521,014 SNPs.

\subsection{Image preprocessing} \label{procmodel}

The T1-weighted baseline MRI scans were downloaded from the Alzheimer’s Disease Neuroimaging Initiative (ADNI) database (n=543). A detailed description of the MRI acquisition protocols can be found on the ADNI website (\url{https:// adni.loni.usc.edu/methods/documents/mri-protocols}). The T1-weighted images $\mathbf{v}$ were then segmented into gray matter (GM), white matter (WM) and cerebrospinal fluid (CSF) tissue compartments using Freesurfer (version 6.0), which is freely available for download (\url{http://surfer.nmr.mgh.harvard.edu}), and has been described previously  \cite{dale1999,fischl1999,fischl2012}. A standardized quality control procedure was used to manually identify and correct any errors in the automated tissue segmentation in accordance with FreeSurfer’s troubleshooting guidelines. Subsequently, cortical GM was parcellated into 68 regions using FreeSurfer’s cortical Desikan-Killiany atlas \cite{desikan2006} and 62 regions using Freesurfer’s Desikan-Killiany-Tourville atlas \cite{klein2012}. Subcortical GM was parcellated into 45 regions using Freesurfer’s “aseg” atlas and subcortical WM was parcellated into 70 regions using Freesurfer’s “wmparc” atlas \cite{fischl2002}. For the white matter parcellation (wmparc), optional Freesurfer parameters were used to ensure the entire white matter compartment was parcellated, (not just WM within a fixed default distance from GM), and any T1 hypotensities were labelled as white matter. This was done to ensure that the white matter parcellation included all white matter voxels and was not biased by individual T1 hypointensity burden. For all other parcellations, the default Freesurfer options were used. From these four parcellations, a total of 1860 features were obtained. These features included:

\bigskip

\begin{itemize}

\item The volume, mean, standard deviation, min, max, and range of Freesurfer normalized T1 intensity values for the “aseg” (270 total features) and  “wmparc” (420 total features) atlas parcellations.

\item The number of vertices, surface area, gray matter volume, thickness (mean, standard deviation), curvature (mean, Gaussian), folding index, and curvature index for the Desikan-Killiany-Tourville (558 total features) and Desikan-Killiany (612 total features) atlas parcellations. 
\end{itemize}

These 1860 features form $\mathbf{v}^*$, the processed image. 

\subsection{Prediction model for feature extraction} \label{predmodel}

 We propose a fully-connected NN as a prediction model for this test application. The inputs of our prediction model are the entirety of the features extracted with FreeSurfer described previously, $\mathbf{v^*}$. The output is AD diagnosis, which is a categorical variable for the ADNI1 data base and finally, the second to last layer of this NN are the features we are interested in. 

\bigskip

In this proposed approach, there is great flexibility to build the early stages of the NN, the encoder equivalent. Specifically, we have control over the number of hidden layers and the non-linear activation function. Assuming the response is a $K_L$-class categorical variable, the output of the NN is a $K_L$-dimensional vector $\mathbf{o}$ where $o_{n,k} = \hat{P}(y_n = k)$, the predicted probability that subject $n$ belongs to class $k$. The relation between the second to last layer and the output layer can be thought of as the one established between predictors and output in a multi-class logistic regression. To do so, we take $K_L$ linear combinations of the $K_{L-1}$ inputs $\mathbf{x}_{L-1}$, so that $\mathbf{o^*} = B\mathbf{x}_{L-1}$, where $B$ is a $K_L \times K_{L-1}$ matrix of coefficients. Then, as activation function, we apply, element-wise, the softmax function to make sure the values are positive and sum to one:  $o_{j} = \frac{\exp(o_j^*)}{\sum_{k=1}^K \exp(o_k^*)}$. 

\bigskip

The model is trained in a similar fashion to a multi-case logistic regression. We minimize the negative log likelihood loss $NLLL(\mathbf{o},\mathbf{y})= \sum_{n=1}^N nlll_n$  where $nlll_n = - \sum_{k=1}^{K_L} \log(o_{n,k}) 1(y_n=k)$. This is essentially the equivalent of maximizing the log likelihood of a multinomial distribution. Thus, one could think of the features extracted $\mathbf{x}_{L-1}$ to effectively be \textit{one logistic regression away} from the disease response. However, these features are constructed from data-driven functions built from the input and one could think of the logistic regression step as the decoder. 

\bigskip

We use the Python language and the Panda package \cite{mckinney2010} to import and manipulate the data set. The feature extraction is entirely done using Python. We use the Pytorch package \cite{Paszke2019} to define and train the NNC. Our NN is a single hidden layer NN with 35 hidden nodes trained with the Adagrad \cite{Duchi2011} optimizer. Finally, in order to train the NNC to distinguish AD from NC patients and thus to extract features related with the difference between those two groups, we only keep NC and AD during the training of the NNC, thus excluding MCI patients. In other words, we train the NNC on a cohort of 287 subjects (145 NC and 142 AD). 

\bigskip

Almost all of the parameters, such as the number of hidden layers (1), the optimizer (Adagrad), the learning rate (0.01), the learning decay (0) and the number of epochs (350) were selected using cross-validation with the exception of the number of neurons in the hidden layer. We have initially set the number of neurons in the second to last layer to 56 as we wanted to design our model to extract the same number of features as in previous articles \cite{greenlaw2017}, \cite{shen2010whole}, and \cite{szefer2017multivariate}. However, reducing its number of neurons to 35 did not decrease the accuracy, so our final set of automatically-extracted features has 35 brain features.

\subsection{Inference model} \label{infemodel}





The SNPs dimension contrasts with its small fraction expected to be related to the imaging phenotypes. SNPs are connected to traits through various pathways and multiple SNPs on one gene often jointly carry out genetic functionalities. Therefore, it is desirable to develop a model to exploit the group structure
of SNPs.

\bigskip

Wang et al. \cite{Wang2012} developed Group-Sparse Multi-task Regression and Feature Selection (G-SMuRFS) to perform simultaneous estimation and SNP selection across phenotypes. Consider matrices as boldface uppercase letters and vectors as boldface lowercase letters. Given the SNP data of the ADNI participants as
$\{\boldsymbol{g}_1,...,\boldsymbol{g}_n\}\subseteq \mathbb{R}^{S}$, where $n$ is the number of participants (sample size), $S$ is the number of SNPs (feature dimensionality), $\boldsymbol{G}=[\boldsymbol{g}_1,...,\boldsymbol{g}_n]$, and the imaging phenotypes as $\{\boldsymbol{x}_1,...,\boldsymbol{x}_n\}\subseteq \mathbb{R}^{C}$, $C$ the number of imaging phenotypes, $\boldsymbol{X}=[\boldsymbol{x}_1,...,\boldsymbol{x}_n]$, $\boldsymbol{W}$ being a $S \times C$ matrix of regression coefficients, where the entry $w_{ij}$ of the weight matrix $\boldsymbol{W}$ measures the relative importance of the $i$-th SNP in predicting the response of the $j$-th imaging phenotype, the matrix algebraic mathematical formulation of the regression is:


\[ \underset{\rm \boldsymbol{W}}{\rm min} ||\boldsymbol{W}^{T}\boldsymbol{G} - \boldsymbol{X}||_F^{2} + \gamma_1 || \boldsymbol{W}||_{Gr_{2,1}} + \gamma_2 || \boldsymbol{W}||_{2,1} \]
where $||.||_{Gr_{2,1}}$ is the group $l_{2,1}$-norm, devised by Wang et al. \cite{Wang2012}. We recapitulate this norm definition: consider that the SNPs, are partitioned into $Q$ groups $\Pi = {\{\pi_q\}}^Q_{q=1}$, such that, the $i$-th row of $\boldsymbol{W}$,  ${\{\boldsymbol{w^i}\}}^{m_q}_{i=1} \in \pi_q$ are genetically linked, $m_q$ being the number of SNPs in $\pi_q$. Denote $\boldsymbol{W}=[\boldsymbol{W}^1 ... \boldsymbol{W}^Q]^T$, $\boldsymbol{W^q} \in \mathbb{R}^{m_q \times c} (1\leq q \leq Q)$, then the group $l_{2,1}$-norm can be both defined as

\[ ||W||_{G_{2,1}} = \sum_{q=1}^{Q} \sqrt{\sum_{i\in\pi_q}\sum_{j=1}^{c} w^2_{ij}} = \sum_{q=1}^{Q}|| \boldsymbol{W^q}||_F\]

While producing sparse point estimates of regression coefficients, the G-SMuRFS lacked standard error computation. Boot-strapping standard error computations were demonstrated to preform poorly when the true value of the coefficient is zero \cite{Kyung2010PenalizedRS}, so an equivalent hierarchical Bayesian model was developed in \cite{greenlaw2017}. The hierarchical model takes the form 
\begin{equation*}
\label{model - level 1}
\mathbf{x}_\ell |\mathbf{W},\sigma^2  \distas{ind} MVN_c (\vW^{T} \mathbf{g}_\ell \: , \: \sigma^2I_c),   \ell=1, \dots, n, 
\end{equation*}
with the coefficients corresponding to different genes assumed conditionally independent
$\mathbf{W}^{(q)}| \lambda_{1}^{2}, \lambda_{2}^{2}, \sigma^2  \distas{ind} p(\mathbf{W}^{(q)}|  \lambda_{1}^{2}, \lambda_{2}^{2}, \sigma^2) \hspace{8pt} q=1,\dots,Q, $
and with the prior distribution for each $\mathbf{W}^{(q)}$ having a density function that is based on a product of multivariate Laplace kernels
\begin{equation*}
\label{PML}
\begin{split}
p(\mathbf{W}^{(q)} |  \lambda_{1}^{2}, \lambda_{2}^{2}, \sigma^2) \propto \exp \left\lbrace - \frac{\lambda_{1}}{\sigma} \sqrt{ \sum_{i \in \pi_q} \sum_{j=1}^c w_{ij}^2 } \right\rbrace  \prod_{i \in \pi_q} \exp \left\lbrace -\frac{\lambda_{2}}{\sigma} \sqrt{\sum_{j=1}^c w_{ij}^2 } \right\rbrace.
\end{split}
\end{equation*}
This product Laplace density can be expressed as a Gaussian scale mixture which allows for the implementation of Bayesian inference using a standard Gibbs sampling algorithm. The algorithm is implemented in the R package \emph{bgsmtr}, \url{https://cran.r-project.org/web/packages/bgsmtr/bgsmtr.pdf} which is available for download on the Comprehensive R Archive Network (CRAN). The selection of tuning parameters $\lambda_{1}$, $\lambda_{2}$ in this model requires cross-validation.

\bigskip

This model serves as our primary inference model in this test application and we refer to this model by the name of its associated package, BGSMTR.

\subsection{Results}

To visualize the proposed set of NN-extracted features in comparison to the features selected based on standard ROIs, we compute a 2-dimensional embedding for both sets of features using a $t$-distributed stochastic neighbor embedding ($t$-SNE) as proposed in \cite{van2008}, a $t$-distributed variant of the original SNE proposed in \cite{hinton2002}. Different from PCA  that finds a linear representation capturing as much variability as possible, SNEs try to identify a low-dimensional representation to optimally preserve a neighborhood identity \cite{hinton2002}. A neighborhood-preserving embedding is especially interesting here as the features are extracted to carry information about the disease status of the patient.

\bigskip

\begin{figure}[h]
\centering
   \includegraphics[width=0.75\linewidth]{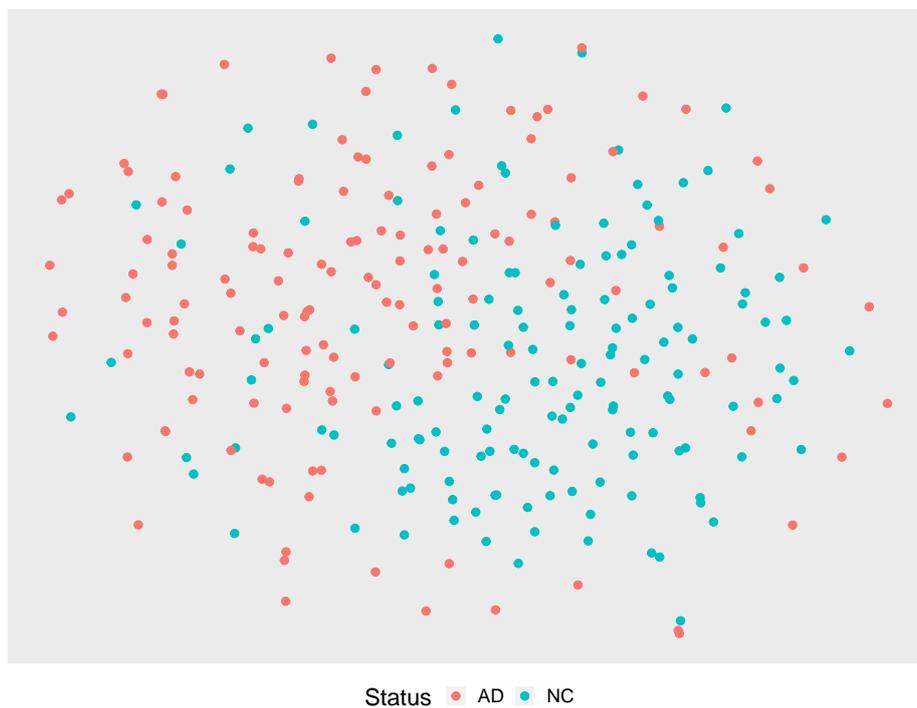}
   \caption{2-dimensional embedding of the NN-extracted features.}
   \label{fig:Ng1} 
\end{figure}

\begin{figure}[H]
\centering
   \includegraphics[width=0.75\linewidth]{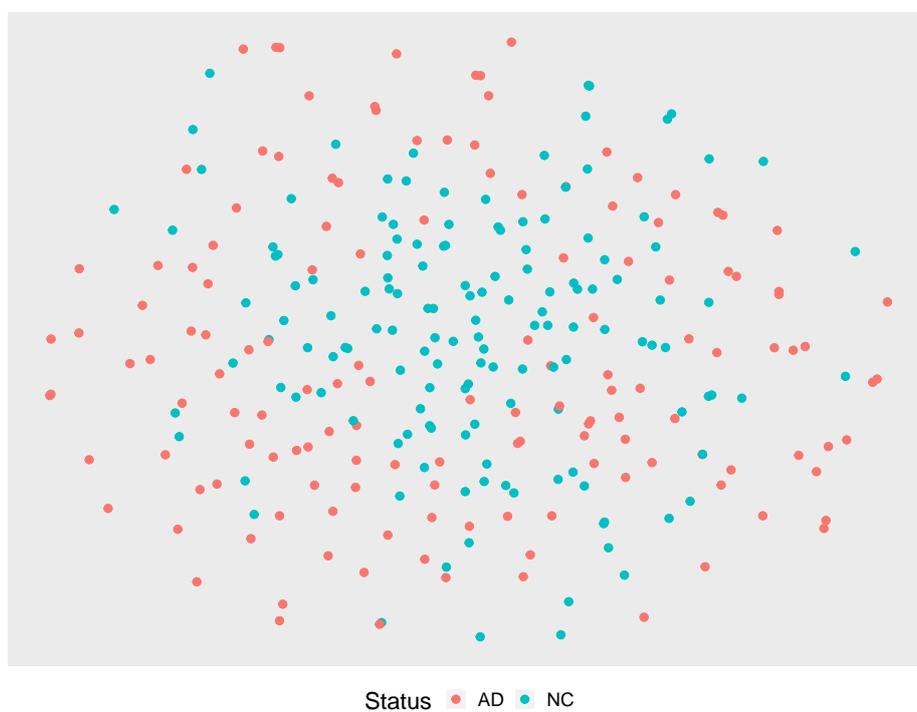}
   \caption{2-dimensional embedding of the expert-selected features.}
   \label{fig:Ng2}
\end{figure}

Figure \ref{fig:Ng1} and \ref{fig:Ng2} contain the embeddings of the training cohort containing strictly the NC and AD patients. When inspecting the figures, it is easier to linearly separate the patients by their status using the NN-extracted features. A randomly selected neighborhood in Figure \ref{fig:Ng1} is more likely to have a high concentration of one class compared to a randomly selected neighborhood in Figure \ref{fig:Ng2}.

\bigskip

\bigskip

Based on the assumption that neuroimaging features that can accurately predict disease status are more closely related to the disease, we compare the accuracy performances of logistic regression models that take our features as inputs, and the model that utilizes features as in \cite{greenlaw2017}, \cite{shen2010whole}, and \cite{szefer2017multivariate}. For that purpose, we proceed with 50 repetitions of random sub-sampling validation: randomly dividing the data set into a training set and a test set. The training set contains 200 observations while the 73 other observations are assigned to the test set. Compared to $k$-fold cross-validation, random sub-sampling validation has the benefit of allowing us to fix the size of the training and testing set independently from the number of Monte Carlo samples.

\bigskip

Table \ref{AccTable} shows the results. The model trained using the automatically extracted features not only has a significantly higher accuracy ($p$-value $< 0.0001$) but also has a smaller performance variance across the sub-samples. The better prediction performance suggests that these features are useful for subsequent genetic analysis.

\begin{table}[H]
\begin{center}
\begin{tabular}{ c| c| c}
\hline
 Features & Mean & Standard Dev. \\
 \hline
 Expert & 0.81808 & 0.03552 \\ 
 Automatic & 0.91726 & 0.02340 \\
 \hline
\end{tabular}
 \caption{\label{AccTable} Mean and standard deviation of the accuracy of a logistic regression that separate NC from AD using two different sets of features: the ROI-based features (Expert) and the features automatically extracted by our proposed NN classifier (Automatic).}
 \end{center}
 \end{table}

 To begin the genetic analysis we adjust for subject specific factors by fitting univariate least squares linear regression for every feature (both NN-derived and ROI-based features) onto the age, gender, education level, the APOE genotype and the total intracranial volume. The residuals from each regression are then used as the adjusted imaging response \cite{szefer2017,song2021} in the inference model. 
 
 \bigskip
 
 We then proceed with a two-step process to reduce the number of SNPs selected. First, we reduce the large number of SNPs to a smaller subset of 485 potentially related with AD SNPs \cite{greenlaw2017} based on expert advices. Second, we fit univariate models between every feature and every SNP and keep the top 100 SNPs based on the resulting $p$-values \cite{yin2011sparse}, \cite{li2015multivariate}. We rank the SNPs by their smallest $p$-value, among all models they are included in.

\begin{table}[h]
\resizebox{\textwidth}{!}{%
\centering
\begin{tabular}{l c c||l c c}
\multicolumn{3}{c}{SNPs related to NN-extracted features} &\multicolumn{3}{c}{SNPs related to Expert-extracted features} \\ \hline
\multicolumn{1}{c}{SNP} & Gene & Status & \multicolumn{1}{c}{SNP} & Gene & Status \\ 
\hline
    \textbf{rs12758257}$^{(2)}$ & ECE1 &  & rs17399090 & DAPK1 & known(\citep{greenlaw2017})\\
    rs2243581 & SORCS1 &  & \textbf{rs12758257}$^{(1)}$ & ECE1 &  \\
    \textbf{rs213025}$^{(3)}$ & ECE1 &  & \textbf{rs213025}$^{(3)}$ & ECE1 & \\
    \textbf{rs12756690}$^{(25)}$ & ECE1 &  & \textbf{rs4935775}$^{(48)}$ & SORL1 & \\
    rs6584777 & SORCS1 &  & rs2179179 & NEDD9 & \\
            \hdashline
    rs9368621 & NEDD9 &  & rs475639 & PICALM & \\
    \textbf{rs213028}$^{(20)}$ & ECE1 &  & rs17209374 & SORCS1 & \\
    \textbf{rs9461448}$^{(90)}$ & PGBD1 &  & rs6905101 & NEDD9 & \\
    \textbf{rs11006130}$^{(16)}$ & TFAM & known(\citep{Wang2012}) & rs3026841 & ECE1 & known(\citep{Wang2012})\\
    rs3739784 & DAPK1 &  & \textbf{rs2276346}$^{(56)}$ & SORL1 & \\
    \hdashline
    rs7897726 & SORCS1  &  & \textbf{rs212531}$^{(41)}$ & ECE1 & \\
    \textbf{rs12001404}$^{(28)}$ & DAPK1 &  & rs17367504 & MTHFR & \\
    \textbf{rs3128521}$^{(27)}$ & DAPK1 &  & \textbf{rs9468690}$^{(50)}$ & NEDD9 & \\
    rs2450129 & GAB2 &  & rs666682 & PICALM & \\
    \textbf{rs12378686}$^{(43)}$  & DAPK1 &  & rs2064112 & NEDD9 & \\
        \hdashline
    rs1318241 & GAB2 &  & \textbf{rs11006130}$^{(9)}$ & TFAM & \\
    \textbf{rs1114188}$^{(47)}$ & DAPK1 &  & \textbf{rs11218301}$^{(29)}$ & SORL1 & \\
    rs11601559 & SORL1 &  & \textbf{rs3118846}$^{(64)}$ & DAPK1 & known(\citep{greenlaw2017}) \\
    rs731600 & GAB2 &  & \textbf{rs3781827}$^{(45)}$ & SORL1 & \\
    rs1893447 & GAB2 &  & \textbf{rs213028}$^{(7)}$ & ECE1 & \\
    \hline
\end{tabular}}
\caption{Second screening results: top 20 SNPs using simple linear regression (univariate regression) with NN-extracted features and expert-extracted features, respectively. SNPs in bold are found related to both sets of features, the superscripted number identifies the rank of the SNPs when using the other set of features.}
\label{tab:20SNPs_comparison_SLR}
\end{table}

\bigskip

Table \ref{tab:20SNPs_comparison_SLR} contains the top 20 SNPs extracted using univariate regression as explained above.
Status, novel or known, is checked against two previous publications, \cite{Wang2021} and \cite{greenlaw2017}. By comparing the SNPs associated with both sets of features, we first notice the top 3 SNPs are quite similar and that overall many SNPs belong to both groups. Additionally, we notice that the genes are also quite similar between the two sets of screened SNPs. However, we also identified multiple SNPs that were not identified using the original, expert-based, features. The possibility of identifying additional SNPs based on features that are more predictive of disease is the potential added-value of the proposed approach. Thus the NN derived features can be used alongside more standard ROI-based features. These novel SNPs could simply be carrying a gene specific signature but this is also a reason why we rely on a multivariate regression model to determine the final set of SNPs.

\bigskip
 
For this reason, we follow with a subsequent multivariate regression that will better allow us to distinguish between association with the features and confounding SNPs. We use the 100 screened SNPs as predictors in our inference model, the BGSMTR model described earlier. 

\begin{table}[h]
\centering
\begin{tabular}{l c c c c}
\multicolumn{1}{c}{\multirow{2}{*}{SNP}} & \multirow{2}{*}{Gene} & \multirow{2}{*}{Chromosome} & \multirow{2}{*}{Status} & No. of\\ 
& & & & related features\\
\hline
    rs2243581 & SORCS1 & 10 & & 1 \\
    rs1699105 & SORL1 & 11 & known(\citep{greenlaw2017}) & 4 \\
    rs6511720 & LDLR & 19 & & 15 \\
    rs6457200 & NEDD9 & 6 & & 8 \\
    rs11006130 & TFAM & 10 & known(\citep{Wang2012}) & 3 \\
    \hdashline
    rs2025935 & CR1 & 1 & known(\citep{greenlaw2017, Wang2012}) & 1 \\
   rs1568400 & THRA & 17 & known(\citep{greenlaw2017}) & 1 \\
    rs3785817 & GRN & 17 & & 11 \\
    rs3026845 & ECE1 & 1 & & 1 \\
    rs12209631 & NEDD9 & 6 & known(\citep{greenlaw2017}) & 5 \\
    \hdashline
    rs2418828 & SORCS1 & 10 & & 1 \\
    rs3118846 & DAPK1 & 9 & known(\citep{greenlaw2017}) & 1 \\
    rs213037 & ECE1 & 1 & & 1 \\
    rs1801131 & MTHFR & 1 & & 3 \\
    rs9368621 & NEDD9 & 6 & & 1 \\
    \hdashline
    rs3793647 & DAPK1 & 9 & & 2 \\
    rs17014873 & BIN1 & 2 & & 1 \\
    rs12758257 & ECE1 & 1 & & 1 \\
    rs762484 & TF & 1 & & 1 \\
    rs2182335 & NEDD9 & 6 & known(\citep{Wang2012}) & 1 \\
    \hline
\end{tabular}
\caption{BGSMTR results: top 20 SNPs related to NN-extracted features with the highest standard score. The last column counts the number of NN-extracted features for which a 95\% credible interval excluded zero.}
\label{tab:25SNPs_comparison_Bayesian}
\end{table}

\bigskip

Table \ref{tab:25SNPs_comparison_Bayesian} contains the top 20 SNPs ranked by the posterior standard score: the posterior mean divided by the posterior standard deviation. In this table we see again a mix of novel and known SNPs and once again, the status, novel or known, is checked against two previous publications, \cite{Wang2021, greenlaw2017}. Among other, identifying the association with AD through MRI features of SNPs rs1699105, rs1699105, rs2025935 and rs12209631 to name a few is consistent with previous publications \cite{Wang2021,greenlaw2017}. On the flip side, if we only discover known SNPs then there is little advantage to our approach. The SNP rs6511720 is ranked very high on the list and was associated with 15 features (according to 95\% credible intervals, not accounting for multiple testing). The SNPs rs6457200, rs2243581  and rs3785817 are also ranked high and/or are related with multiple features.

\bigskip

The results above provide a strong argument in favour the proposed pipeline; the features extracted are not only better at predicting the neurological disease of interest but also allowing the identification of different SNPs. Furthermore, the extraction process is data-driven and requires no expert advice, outside of the diagnostic.

\section{Discussion} \label{Discussion}

Given the results of the previous section, we argue in favor of using automatic feature extraction in addition to ROI or voxelwise features to find signal potentially novel SNPs that may not be detected when using ROIs or voxels alone. Our focus here is to identify SNPs related with MRI in a manner that is predictive of disease and obtain confidence intervals and posterior distributions. Integrating machine learning approaches within imaging genetics studies is of potential use as demonstrated in our analysis. 

\bigskip

Additionally, our work demonstrates the use of different objective functions to extract features and reduce the dimension of large  observations, such as neuroimages. Instead of using unsupervised models, we are able to direct the feature extraction towards a variable of interest, in our case the diagnostic variable. However, with gradient-based models, such as NN, we can design many other objective functions and tailor the feature extraction process for problem specific needs. This idea can be applied in various ways when we analyse neuroimages, and we recommand considering a large collection of objective functions that are data-driven when extracting features instead of strictly relying on expert-advice. 

\bigskip

We choose to use a NN for feature extraction, this comes with strengths and weaknesses. Because our goal is to do inference at the SNP level we agreed to lose interpretability on the neuroimage feature level, this is usually considered a weakness of blackbox models such as NNs. In counterparts, this allows us to get complicated features that are functions of the complete processed images and the use of classification models ensure that those features are indeed related to AD. We acknowledge a loss in the interpretability of regression coefficients. Nevertheless we believe the automatic feature extraction approach provides added value when used alongside studies that are conducted at either the ROI or voxelwise level. It requires no external expertise for feature selection. The features are built considering disease prediction through nonlinear representations of neuroimaging.

\bigskip

 One advantage of the procedure we propose is that we can easily improve on each of the three pieces of the pipeline separately. One avenue for future work is to review how to process the images. In this first implementation of our proposed pipeline we use the well established FreeSurfer software to obtain volumetric and cortical thickness statistics from the MRI scans. However, we believe the fact our features were automatically extracted in a data-driven way was the reasons why our features have higher predictive power. Thus, it seems reasonable to extend that principle to image processing and also try to automatically process the images in a data-driven way. For instance, a common NNC for images is the Convolutional Neural Networks (CNNs) \cite{lecun1989,lecun2015,goodfellow2016}.  Using a CNN taking as input the 3-dimensional brain scan images and training this model to predict the diagnosis would be of potentially great  value for further investigation. The convolutional layers replace some of the image processing steps and the lower-level layers act as the feature extractor. However, some processing, mostly registration, would still be required. Another interesting approach to explore is to use an AE to reduce the dimension of the images in an unsupervised manner first. Different AEs can be trained for each brain regions separately and it allows the number of features extracted per region to vary. This allows the collection of AEs to extract more features from regions with higher variability or from regions with more predictive power. 

\bigskip

Finally, the last step of our pipeline involves an inference step using a multivariate Bayesian group sparse regression. There is scope for generalizing this step to account for model uncertainty where the Bayesian model used is included within a collection of different models (e.g., 
\cite{batmanghelich2013joint}  \cite{batmanghelich2016probabilistic} \cite{stingo2013integrative}   \cite{zhu2014bayesian}  \cite{kundu2016semiparametric}) and then Bayesian model averaging is used for inference at the SNP level while accounting for model uncertainty.

\section*{Acknowledgements}

The authors would like to acknowledge the financial support of the Canadian Statistical Sciences Institute (CANSSI), the Alzheimer Society Research Program (ASRP), the National Health Institute (NIH) and the Natural Sciences and Engineering Research Council of Canada (NSERC). 


\bigskip

Data collection and sharing for this project was funded by the Alzheimer's Disease Neuroimaging Initiative
(ADNI) (National Institutes of Health Grant U01 AG024904) and DOD ADNI (Department of Defense award
number W81XWH-12-2-0012). ADNI is funded by the National Institute on Aging, the National Institute of
Biomedical Imaging and Bioengineering, and through generous contributions from the following: AbbVie,
Alzheimer’s Association; Alzheimer’s Drug Discovery Foundation; Araclon Biotech; BioClinica, Inc.; Biogen;
Bristol-Myers Squibb Company; CereSpir, Inc.; Cogstate; Eisai Inc.; Elan Pharmaceuticals, Inc.; Eli Lilly and
Company; EuroImmun; F. Hoffmann-La Roche Ltd and its affiliated company Genentech, Inc.; Fujirebio; GE
Healthcare; IXICO Ltd.; Janssen Alzheimer Immunotherapy Research \& Development, LLC.; Johnson \&
Johnson Pharmaceutical Research \& Development LLC.; Lumosity; Lundbeck; Merck \& Co., Inc.; Meso
Scale Diagnostics, LLC.; NeuroRx Research; Neurotrack Technologies; Novartis Pharmaceuticals
Corporation; Pfizer Inc.; Piramal Imaging; Servier; Takeda Pharmaceutical Company; and Transition
Therapeutics. The Canadian Institutes of Health Research is providing funds to support ADNI clinical sites
in Canada. Private sector contributions are facilitated by the Foundation for the National Institutes of Health
(www.fnih.org). The grantee organization is the Northern California Institute for Research and Education,
and the study is coordinated by the Alzheimer’s Therapeutic Research Institute at the University of Southern
California. ADNI data are disseminated by the Laboratory for Neuro Imaging at the University of Southern
California.

\pagebreak
\bibliography{mybibfile}

\end{document}